\documentclass[letterpaper, 11pt]{article}
\pdfoutput=1

\usepackage[paperwidth=8.5 in, paperheight=11 in, 
	left=1 in, right=1 in, bottom=1 in, top=1 in]{geometry}

\usepackage[]{authblk}

\usepackage{amsmath}
\usepackage{amssymb}

\usepackage{graphicx}
\DeclareGraphicsExtensions{.pdf,.png,.jpg}

\usepackage[backend=biber, style=ieee, autocite=plain,sorting=none]{biblatex}
\newcommand{\vect}[1]{\boldsymbol{#1}}

\begin{document}
\title{Quantifying the Structure of Disordered Materials}
\author[1]{Thomas J. Hardin}
\author[1]{Michael Chandross}
\author[2]{Rahul Meena}
\author[3]{Spencer Fajardo}
\author[2]{Dimitris Giovanis}
\author[4,5]{Yannis Kevrekidis}
\author[3,6,7,8]{Michael Falk}
\author[2,8]{Michael Shields}

\affil[1]{Computational Materials and Data Science Department, Sandia National Laboratories}
\affil[2]{Department of Civil Engineering, Johns Hopkins University}
\affil[3]{Department of Materials Science and Engineering, Johns Hopkins University}
\affil[4]{Department of Applied Mathematics and Statistics, Johns Hopkins University}
\affil[5]{Department of Chemical and Biomolecular Engineering, Johns Hopkins University}
\affil[6]{Department of Mechanical Engineering, Johns Hopkins University}
\affil[7]{Department of Physics and Astronomy, Johns Hopkins University}
\affil[8]{Hopkins Extreme Materials Institute, Johns Hopkins University}
\maketitle

\begin{abstract}
Durable interest in developing a framework for the detailed structure of glassy materials has produced numerous structural descriptors that trade off between general applicability and interpretability. However, none approach the combination of simplicity and wide-ranging predictive power of the lattice-grain-defect framework for crystalline materials. Working from the hypothesis that the local atomic environments of a glassy material are constrained by enthalpy minimization to a low-dimensional manifold in atomic coordinate space, we develop a novel generalized distance function, the Gaussian Integral Inner Product (GIIP) distance, in connection with agglomerative clustering and diffusion maps, to parameterize that manifold. Applying this approach to a two-dimensional model crystal and a three-dimensional binary model metallic glass results in parameters interpretable as coordination number, composition, volumetric strain, and local symmetry. In particular, we show that a more slowly quenched glass has a higher degree of local tetrahedral symmetry at the expense of cyclic symmetry. While these descriptors require post-hoc interpretation, they minimize bias rooted in crystalline materials science and illuminate a range of structural trends that might otherwise be missed.
\end{abstract}

\section*{Main}
The ability to interrogate, understand and modify the microscopic \emph{structure} of materials distinguishes modern materials scientists from artisans of old. Starting in the 1800s \autocite{sorby1886application} and developing to this day, the capacity to pinpoint, classify, and count dislocations, vacancies, interstitials, grain boundaries, grain sizes, and other deviations from tabulated lattices revolutionized crystalline materials science \autocite{azevedo2020history}, constituting one corner of the materials science tetrahedron \autocite{national1989materials}. The success of this structure-centric approach has sparked lasting interest in developing a similarly intuitive and useful framework for the structure of non-crystalline materials, which lack crystal symmetry but which nonetheless feature structural patterns with first-order effects on properties \autocite{pan2018extreme, luan2022high}.

While the lattice-grain-defect framework produces interpretable and widely applicable descriptions of crystalline materials, efforts to describe glassy structure have required tradeoffs between interpretability and generality. At the most-general extreme are atomic coordinates, which are high-dimensional and (in a glass) can be bewilderingly complex. At the other (interpretation-friendly) extreme is the simple question, ``is there a crystal lattice,'' which is of interest but which is inadequate to predict the variety of behavior observed within families of chemically similar glassy materials \autocite{pan2018extreme, luan2022high, ganapathi2021structure}. Since long-range order is absent in glassy materials, descriptors between these extremes have focused on short- and medium-range structure. 

Physically-motivated scalar descriptors such as free volume \autocite{spaepen1977microscopic, spaepen2006homogeneous} and flexibility volume \autocite{ding2016universal, fan2017correlating} emphasize interpretability and performance on a focused set of problems (such as predicting ease of plastic deformation) \autocite{richard2020predicting}. Machine-learned scalar descriptors \autocite{fan2020machine} have made some progress towards generalizing to multiple problems. In covalent glasses, topological constraint theory correlates coordination number with a range of mechanical and chemical properties \autocite{pignatelli2016topological, bauchy2020topological}. However, scalar descriptors are necessarily lossy, discarding information that might be illustrative or useful in some other context.

Other descriptors aspire to much broader usefulness, attempting to distill the important structural features while discarding redundant or noisy degrees of freedom without passing judgment on any given feature's usefulness. Ideally, this would result in a description that reflects much of the complexity of a glass but is lower-dimensional and more interpretable than raw atomic coordinates. Intermediately-lossy descriptors in this category include the statistics of rings in a covalent glass \autocite{zhou2021experimental} and the radial distribution function \autocite{mu2021unveiling}. More granular descriptors include the Z-clusters formalism \autocite{ma2015tuning} and related efficient packing theory \autocite{miracle2004structural, sheng2006atomic} for metallic glasses, which categorize the first-nearest-neighbor Voronoi polyhedra in a sample and rationalize the geometric frustration in packing those polyhedra together into a solid, respectively.

Here we present a data-driven approach to describing the short-range structure of glassy materials that is more general and less lossy than the structural descriptors mentioned above, but more interpretable than raw atomic coordinates. Our approach is based on the notion that a local atomic environment (LAE) consisting of $n$ atoms can be thought of as a point in a $4n$-dimensional space (encoding three spatial dimensions and one chemical dimension for each atom). The ensemble of LAEs in a glassy sample, then, comprise a point cloud in that high-dimensional space. We hypothesize that enthalpy minimization loosely constrains those points onto an energetically-favorable lower-dimensional manifold in $4n$-space, while kinetics and entropy spread them out on that manifold. In this framing, the problem of finding a complete and parsimonious set of glassy structural descriptors is equivalent to parameterizing this manifold \autocite{reinhart2017multi, freitas2020uncovering}, through the use of manifold learning and nonlinear dimensionality reduction.

Our strategy is to:
\begin{enumerate}
\item sample LAEs from a material (each a point on the material's local structural manifold in $3n+n$-space),
\item quantify the difference between each pair of sampled LAEs (forming a square matrix of generalized distances between those points),
\item use that matrix of distances as input for agglomerative clustering \autocite{lance1967general} and diffusion maps \autocite{coifman2006diffusion} (well-established dimensionality reduction techniques), and
\item interpret the unlabeled classes and diffusion coordinates in terms of physical quantities.
\end{enumerate}
Specifically, we used a novel generalized distance function, the Gaussian Integral Inner Product (GIIP) distance, to compare LAEs in step 2 of this strategy. The GIIP distance (inspired by the Smooth Overlap of Atomic Potentials or SOAP formalism \autocite{bartok2013representing}), described in detail in the Methods section, is continuous and smooth with respect to atomic perturbation and insensitive to the orientations of the LAEs. 
The resulting structural description emphasizes generality (being application-agnostic), minimizes bias from human inputs, and illuminates nontrivial relationships between structural configurations in the glass.

\section*{Results}
We applied our strategy to two samples. Our first sample is an easily understood two-dimensional model crystal with defects, which we use to illustrate both our manifold learning strategy and an approach to interpreting the data-mined parameters. Then we considered a far more complex three-dimensional model binary metallic glass, which was the motivating use case for the strategy.

In each case, to improve the computational tractability of our strategy, we partitioned the sampled LAEs into \textit{training} and \textit{extension} sets. Rather than calculating the GIIP distance between every pair of LAEs sampled, we calculated the GIIP distance only between pairs of members of the training set, and between members of the training and extension sets. This serves to reduce the number of GIIP distances calculated from O($n_{\text{total}}^2$) to O($n_{\text{train}} \times n_{\text{total}}$). We performed agglomerative clustering first solely on the training set to establish classes of LAEs, and then assigned LAEs in the extension set to the class to which the nearest LAE in the training set belonged. 
In a similar vein, we calculated the diffusion map first only on the LAEs in the training set to define a set of diffusion coordinates for the system, and then used the Nystr\"om extension \autocite{Nystrom_Extension} to map the extension-set LAEs into those already-defined diffusion coordinates. This approach enabled us to include many more atoms in our analysis than computational limitations would permit otherwise.

The GIIP distance is constructed so that the distances between configurations have units of ``atoms of difference.'' For example, if LAE 1 is identical to LAE 2 with the exception of LAE 2 having a single void where LAE 1 has an atom, the GIIP distance between LAEs 1 and 2 would be approximately one. In binary and higher-order systems, an atom with chemical species A aligning with an atom with chemical species B would produce a GIIP distance of approximately two. Minor atomic misalignments between similar LAEs would generally produce a GIIP distance between zero and one.

\subsection*{Two-dimensional crystal}
\begin{figure}
\centering
\includegraphics[width=\textwidth]{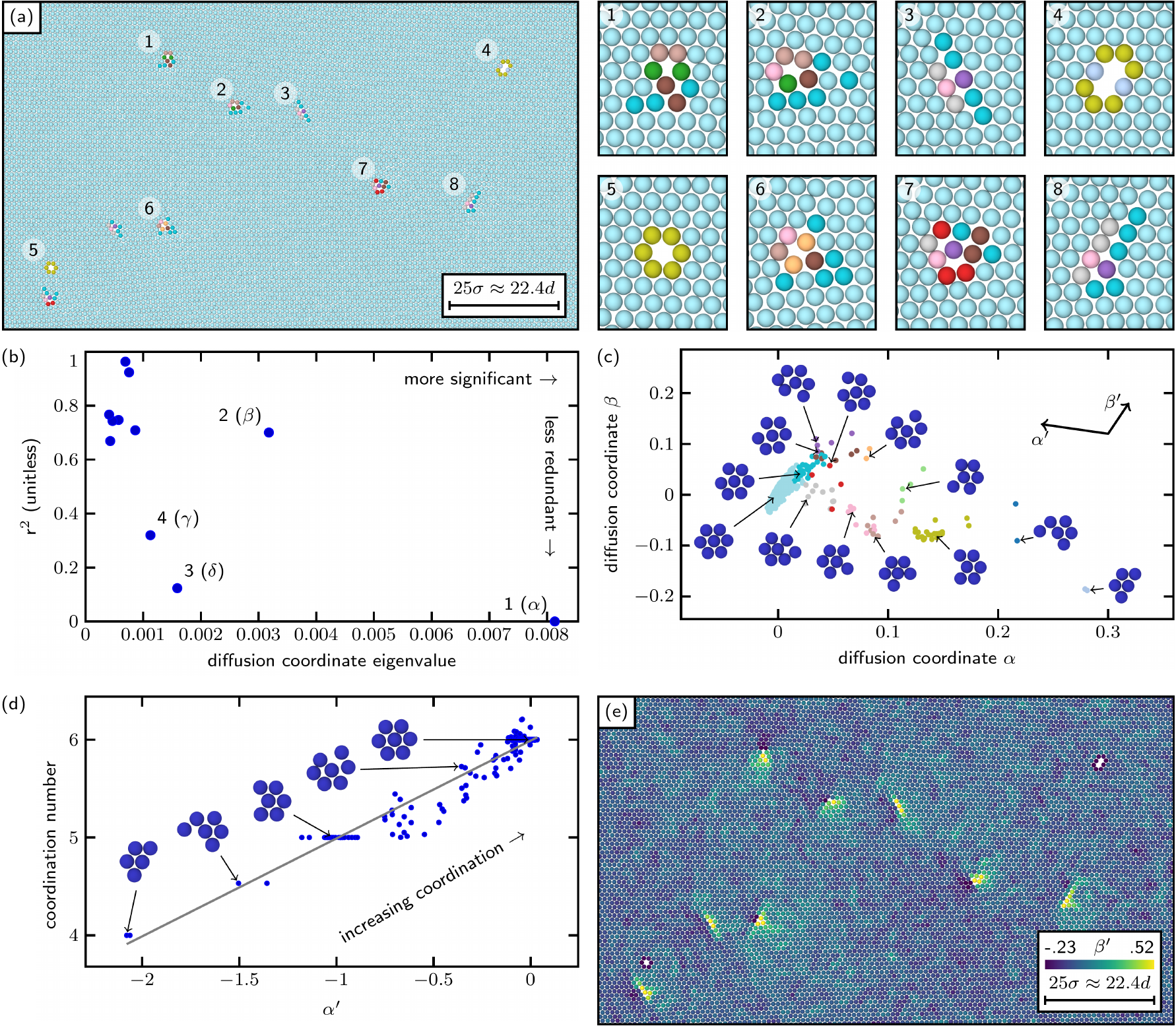}
\caption{(a) rendering of the two-dimensional crystal colored by membership in thirteen classes. (b) plot of the first ten diffusion coordinates in terms of significance and redundancy as described in the text, showing clear separation between the first four diffusion coordinates and the remainder. (c) the sample plotted in the $\alpha\times\beta$-plane of diffusion space, colored by class as in (a), with exemplar environments shown. Directions $\alpha'$ and $\beta'$ are additionally noted. (d) $\alpha'$ correlates to coordination number with an r$^2$ of 0.93. (e) the atoms from (a) colored by $\beta' $, revealing two-lobed volumetric strain fields as would be expected around dislocation cores.}
\label{fig:xl2}
\end{figure}

We modeled a defective two-dimensional crystal of 19548 atoms with a Lennard-Jones interatomic potential (as described in the Methods section). A portion of the sample is shown in Fig.~\ref{fig:xl2}(a). We identified 15484 atoms away from the sample edges and ordered them in terms of potential energy. We selected a training set consisting of first-nearest-neighborhood LAEs centered on the 100 atoms with the lowest potential energy, and the 4900 atoms with the highest potential energy, and designated the remaining 15484-5000=10484 LAEs as the extension set. As described above, we evaluated the GIIP distances between the sampled LAEs and, applying agglomerative clustering to the raw GIIP distances, found that the LAEs  could be partitioned into 13 classes with no two members of any class being more than one atom distant from each other (by GIIP distance). The atoms in Fig.~\ref{fig:xl2}(a) are colored according to membership in these 13 classes. 

We also generated the diffusion coordinates for the first-nearest-neighborhood LAEs in the sample, using the training and extension set approach described above. Each diffusion coordinate was associated with an eigenvalue that expresses the strength of the coordinate's contribution to variation within the data---that is, larger eigenvalues are more important, while smaller eigenvalues can be truncated without losing significant information. We further examined the redundancy of each successive diffusion coordinate $i$, by attempting to predict it in terms of the preceding diffusion coordinates (1 through $i-1$) with polynomial regression. We calculated the coefficient of determination (r$^2$) of the regression problem, where r$^2$ close to one indicates that the information in the diffusion coordinate of interest was already contained in the preceding diffusion coordinates, and r$^2$ closer to zero indicates that the diffusion coordinate contained new information (an analysis based on \autocite{dsilva2018parsimonious}). The eigenvalue and r$^2$ values for this diffusion map are shown in Fig.~\ref{fig:xl2}(b), where the first four diffusion coordinates (labeled $\alpha$, $\beta$, $\delta$, and $\gamma$) are clearly separated. For this example, we take $\alpha$ through $\gamma$ to be our diffusion map.

The LAEs in our sample are plotted in the ($\alpha \times \beta$)-plane of diffusion space in Fig.~\ref{fig:xl2}(c). The LAEs are colored by their 13 agglomerative clustering classes as in Fig.~\ref{fig:xl2}(a), with exemplar LAEs rendered alongside for each class. The numeric values of the diffusion coordinates are entirely arbitrary, as they are simply a parameterization of a manifold, and could be shifted, scaled, or invertibly nonlinearly transformed and remain a valid parameterization of that manifold. The valuable information is in the relative positions of the points in diffusion space,  how they cluster and the trends they reveal when the manifold is ``unrolled'' into simpler coordinates. The axes are notably not labeled in a physically meaningful way as they only reflect trends in the variation of the data; it is our job to inspect, regress, and interpret the meanings of those trends.

On inspection of Fig.~\ref{fig:xl2}(c) we observed that the agglomerative clustering classes generated from the raw GIIP distance matrix are clustered in diffusion space. A tight grouping of near-perfect-crystal LAEs appear near $\alpha=0,\beta=0$ with defective LAEs scattered alongside. The coordination number of the central atom in each LAE varies consistently along the direction marked $\alpha'$, a correlation confirmed in Fig.~\ref{fig:xl2}(d) (noting again the arbitrary nature of the units of $\alpha'$) by fitting a line with an r$^2$ of 0.93. Taking $\beta'$ to be orthogonal to $\alpha'$ in the $\alpha\times\beta$-plane, and using $\beta'$ to color the atoms in Fig.~\ref{fig:xl2}(e) reveals volumetric strain fields in the vicinity of dislocation cores in the sample, where negative $\beta'$ appears in connection with the compressive fields of extra half-planes of atoms, and positive $\beta'$ appears as balancing tensile fields mirrored across the slip plane \autocite{anderson2017theory}.

Having illustrated our approach, rather than invest in further interpretation of a contrived example, we now move on to a far more complicated and interesting system, namely a three-dimensional binary metallic glass.

\subsection*{Three-dimensional binary metallic glass}
\begin{figure}
\centering
\includegraphics[width=\textwidth]{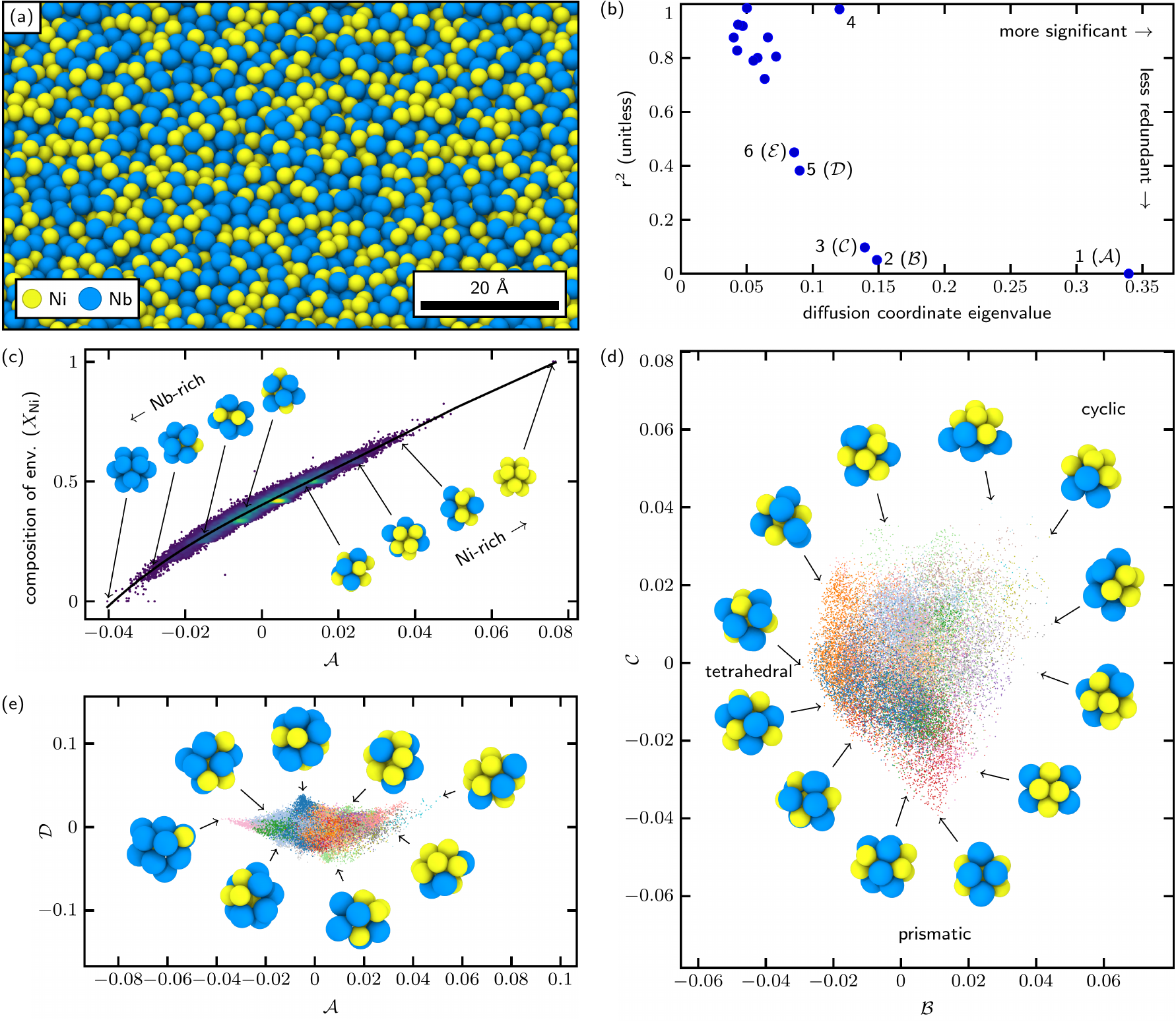}
\caption{(a) rendering of the NiNb glass. (b) diffusion coordinates plotted in terms of significance and redundancy. (c) diffusion coordinate $\mathcal{A}$ correlates to the composition of the first-nearest-neighbor environments (r$^2$=.97), plotted in terms of the fraction of Ni neighbors in the LAE (excluding the central atom). (d) The sampled configurations plotted in the $\mathcal{B}\times\mathcal{C}$-plane of diffusion space, colored by membership in 20 agglomerative clustering classes, with representative configurations shown. The configurations fall in a roughly triangular space with one corner representing tetrahedral symmetry, one corner representing prismatic symmetry, and one corner representing cyclic configurations. (e) Sampled configurations plotted in the $\mathcal{A}\times\mathcal{D}$-plane of diffusion space, colored by class as before. Cusps in the data are noted, suggesting that (for example) certain compositions strongly favor certain structures.}
\label{fig:mg3}
\end{figure}

We used an embedded atom method (EAM) interatomic potential to model a three-dimensional equicompositional NiNb metallic glass. We generated four melt-quench samples of size 13500 atoms each, cooled (respectively) at rates of $1.7\times 10^{13}$, $3.4\times 10^{12}$, $1.7\times 10^{12}$, and $1.7\times 10^{11}$ K/s. A representative rendering of the sample cooled at $1.7\times 10^{11}$ K/s is shown in Fig.~\ref{fig:mg3}(a). The dataset was constructed by pooling the first-nearest-neighborhood LAEs centered on Ni atoms from all four quench-rate samples (27,000 LAEs total). The data was partitioned into a training set consisting of 1,700 LAEs from each quench-rate sample (6,800 training LAEs total) an extension set consisting of the remaining 20,200 LAEs. As before, we evaluated GIIP distances (including species information, so Ni atoms do not align constructively with Nb atoms) and calculated the diffusion map for the sample. An initial attempt at agglomerative clustering found that the variance in LAEs in the sample was so great that constructing a small number of useful classes was impossible; for the purpose of coloring our diffusion space scatter plots, we partitioned the data into twenty classes, but make no claim as to their physical significance.

The diffusion coordinates are plotted in Fig.~\ref{fig:mg3}(b) in terms of significance and redundancy (as described in the previous section) and from this we selected a parsimonious set of five diffusion coordinates (1, 2, 3, 5, and 6) that are well-separated from the other diffusion coordinates. These are labeled $\mathcal{A}$ through $\mathcal{E}$. Fig.~\ref{fig:mg3}(c) shows that diffusion coordinate $\mathcal{A}$ correlates to a high degree with the chemical composition of the local environment.

The LAEs in the sample are plotted in the ($\mathcal{B}\times\mathcal{C}$)-plane of diffusion space in Fig.~\ref{fig:mg3}(d), colored by their membership in the twenty agglomerative clustering classes, with representative LAEs from the convex hull rendered around the perimeter. We find that the LAEs populate a roughly triangular region in this plane, with the corners of representing three extremes of symmetry. At one corner (negative $\mathcal{B}$, small $\mathcal{C}$) we find LAEs with strong tetrahedral symmetry. At another corner (positive $\mathcal{B}$ and $\mathcal{C}$) we find LAEs tending towards cyclic (C$_n$) symmetry---meaning rotational symmetry about a single axis, but few other symmetries. At the last corner (negative $\mathcal{C}$ and small $\mathcal{B}$) we find LAEs with prismatic (particularly D$_{2\text{h}}$) symmetry.

The LAEs are also plotted in the ($\mathcal{A}\times\mathcal{D}$)-plane in Fig.~\ref{fig:mg3}(e), colored as in Fig.~\ref{fig:mg3}(d). We have been unable to develop physical interpretations of $\mathcal{D}$ or $\mathcal{E}$, but note the presence of strong cusps in the envelope of LAEs sampled. These cusps suggest the existence of extrema, either energetic or symmetric, where the number of observed configurations drops to one, meaning that a single configuration is overwhelmingly favored at that point.

Next, we consider the impact of quench rate on the structural state of our model metallic glass.

\subsection*{Effect of quench rate on metallic glass}
\begin{figure}
\centering
\includegraphics[width=\textwidth]{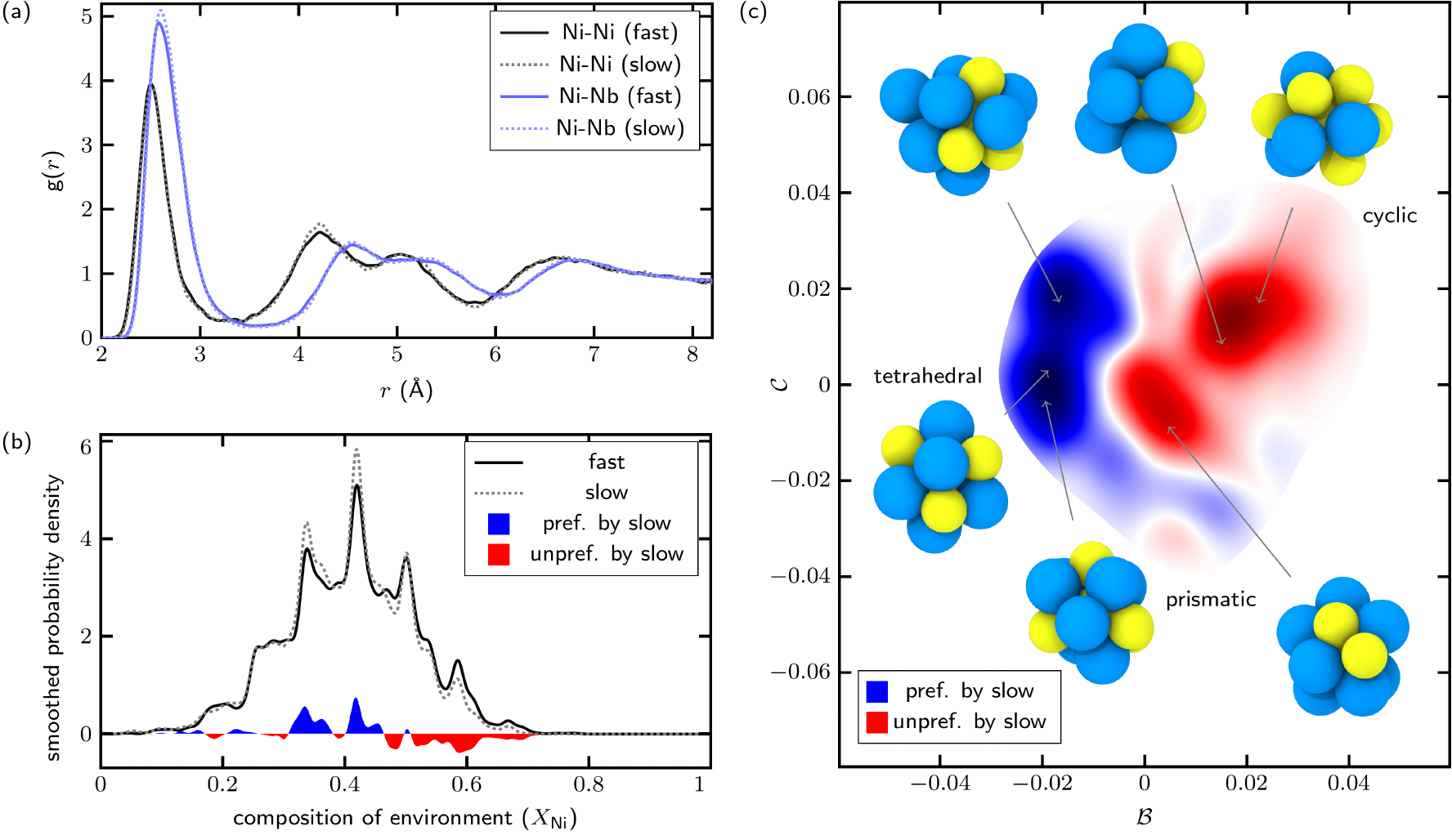}
\caption{(a) Ni-centered partial radial distribution functions for the fastest-quenched and slowest-quenched samples in our metallic glass dataset ($1.7\times 10^{13}$ and $1.7\times 10^{11}$ K/s, respectively). (b) smoothed probability density functions of LAE composition for fastest and slowest quench rates, and difference in blue and red along the x-axis. Blue indicates increased density at the slower quench rate, while red indicates decreased density. (c) difference in smoothed probability density functions for the fastest and slowest quench rates in the $\mathcal{B}\times\mathcal{C}$-plane; red indicates decreased density at the slower quench rate, while blue indicates higher density at the slower quench rate. Slower quench rates seem to favor tetrahedral configurations at the expense of cyclic configurations.}
\label{fig:quench}
\end{figure}

It is well-established that quench rate impacts both the structure and properties of metallic glass \autocite{schawe2019existence}, as slower quench rates allow greater numbers of LAEs to find their way to lower-energy states. We separated out the LAEs from our dataset that come from the fastest-quenched and slowest-quenched samples ($1.7\times 10^{13}$ and $1.7\times 10^{11}$ Kelvin per second, respectively) and compared partial radial distribution functions (Fig.~\ref{fig:quench}(a)) and distributions in diffusion space (Fig.~\ref{fig:quench}(b-c)). Subtle differences between the partial radial distribution functions are observed. Most prominently, the first peak of the Ni-Nb distributions is slightly increased for the slow quench relative to the fast quench, suggesting that slower quenches encourage Nb enrichment of Ni-centered LAEs. In Fig.~\ref{fig:quench}(b) we examine the distributions of chemical compositions of the LAEs (equivalent to diffusion coordinate $\mathcal{A}$) and find that slower quench rates indeed disfavor concentrated Ni-rich LAEs in favor of slightly Nb-rich LAEs, supporting and adding detail to the interpretation of the radial distribution functions. This implies a thermodynamic drive away from segregation as cooling rates increase, but is not particularly detailed in terms of the actual structural transitions occurring in the glass.

To further illuminate these changes, we plot the difference in probability density functions for the two quench rates in the ($\mathcal{B}\times\mathcal{C}$)-plane in Fig.~\ref{fig:quench}(c). In this plot, red indicates regions of diffusion space that are relatively denuded at the slow quench rate, while blue indicates enrichment in the slow sample. We find that slower cooling rates strongly disfavor LAEs with local cyclic symmetry and intermediate configurations, are close to neutral with respect to prismatic LAEs, and strongly favor LAEs with tetrahedral symmetry. This is consistent with the previously-mentioned trend towards ordering of Ni and Nb atoms, which are known to form intermetallic Ni$_3$Nb and Ni$_6$Nb$_7$ phases at equilibrium \autocite{jones2021additive} and illustrates one of the steps that take a sample from liquid to intermetallic states \autocite{sun2016crystal}.

\section*{Discussion}
As described in the introduction, structural descriptors for glass currently in the literature generally involve a trade-off between simplicity (interpretability), and completeness (generalizability). Our approach emphasizes completeness/generalizability and uses data mining to extract trends in the LAEs encountered to serve as a sort of compatibility layer between raw atomic coordinates and a human mind. In the case of the three-dimensional metallic glass, most of the LAEs sampled had about 21 near neighbors within 4 angstroms of the central atom, implying the need for an 84-dimensional space to describe them. The combination of GIIP and diffusion coordinates reduces this to a five-dimensional space that captures the dominant trends in the first-nearest-neighbor structure of the glass. This supports the postulate that the atomic configurations of a metallic glass sit on a (relatively) low-dimensional manifold despite featuring tens of thousands of possible distinct configurations.

Some descriptors in the literature focus on interatomic potential development based on spherical and hyperspherical harmonic expansions (e.g. SOAP \autocite{bartok2013representing}). These descriptors are extremely general, and comparison to GIIP in this particular use case is somewhat artificial, since they are highly optimized for a very different use case from structure identification. We also note that GIIP would be an inefficient method for interatomic potential development. It is worth noting that these descriptors systematically discard structural information (even when truncating their Fourier bases at arbitrarily high numbers of terms) \autocite{pozdnyakov2020incompleteness}, so distance measurements based on these descriptors are inherently lossy. The effects of this will vary on a case-by-case basis but an incomplete generalized distance function is at risk of ``short-circuiting'' \autocite{de2008introduction} the dimensionality reduction methods presented here. However, the speed with which these descriptors can be evaluated might in some cases make this risk worthwhile, particularly for a first-pass structural assessment. We evaluate generalized distances based on the radial distribution function along similar lines.

Descriptors based on diffusion maps and agglomerative clustering are also not lossless, even when the distance function satisfies the identity of indiscernibles. Our approach deals with this by being \textit{tunably} lossy, letting the end user pick the number of discrete classes or diffusion coordinates depending on use case.

It is perhaps more fair to compare our approach to the Z-clusters concept for metallic glasses. In some sense a large set of agglomerative clusters extracted by our approach can be understood as a more granular extension of Z-clusters, from which Z-clusters could be extracted. The Z-clusters approach is inherently discrete in nature, whereas our diffusion maps return a continuous parameterization of structure.  The latter has the potential to be more useful than a discrete parameterization in some settings, depending on success of efforts to interpret the diffusion coordinates.

Comparison of our approach to free or flexibility volume, as well as topological descriptors for silicate glass, highlights its greatest weakness, namely that the computer does not label the coordinates or classes that it returns. Our analyses suggest that it can be possible to identify some coordinates through intuition and trial-and-error; mapping the remaining coordinates to existing descriptors is an ongoing effort, as is developing physical intuition for the coordinates that don't seem to line up with existing ideas. While the opacity of parameters is a current challenge for the application of data-mining to the physical sciences, detection of structural trends, even when difficult to interpret, is ultimately a useful step towards both property prediction and physical insight.

The shortcoming of only partial interpretability is offset by the limited number of user-supplied assumptions underpinning our approach. The GIIP distance simply defines configurations with similar atomic positions as similar and allows for  the determination of trends in the data without bias toward or away from principles rooted in crystalline materials science.

We hypothesize that to some extent the properties of a material will be determined by the distributions of first-nearest-neighbor LAEs in diffusion space---something like a texture map for a polycrystalline metal---but spatial correlations between even fully-characterized LAEs remain difficult to engage. The examples presented in this paper are restricted to first-nearest-neighbor LAEs, but this is not a fundamental limitation of our strategy and it is possible that examination of second- and higher-order neighborhoods will prove illuminating.

In addition to ongoing work to make connections between the data-mined descriptors and existing physical descriptors, we anticipate that diffusion coordinates or their distributions can serve as state variables for constitutive modeling or as microstate definitions in statistical mechanical calculations. A logical next step in this approach is also to data-mine individual kinetic transitions in the glass. Finally, we note that our approach is material-agnostic and has potential applications in the study of liquids, polymers, colloidal systems, granular systems, and other non-crystalline materials. We also note potential synergy with emerging atomic resolution experimental characterization techniques \autocite{yang2021determining}. The rational design of materials requires a fundamental understanding of structure. This methodology is a step towards illuminating glassy structure in a form that is detailed yet interpretable, and a potential step towards a universal coordinate system of machine-learned structural descriptors parameterizing the manifold of physically realizable materials.

\section*{Methods}
\subsection*{Gaussian Integral Inner Product (GIIP) distance}
\begin{figure}
\centering
\includegraphics[scale=1]{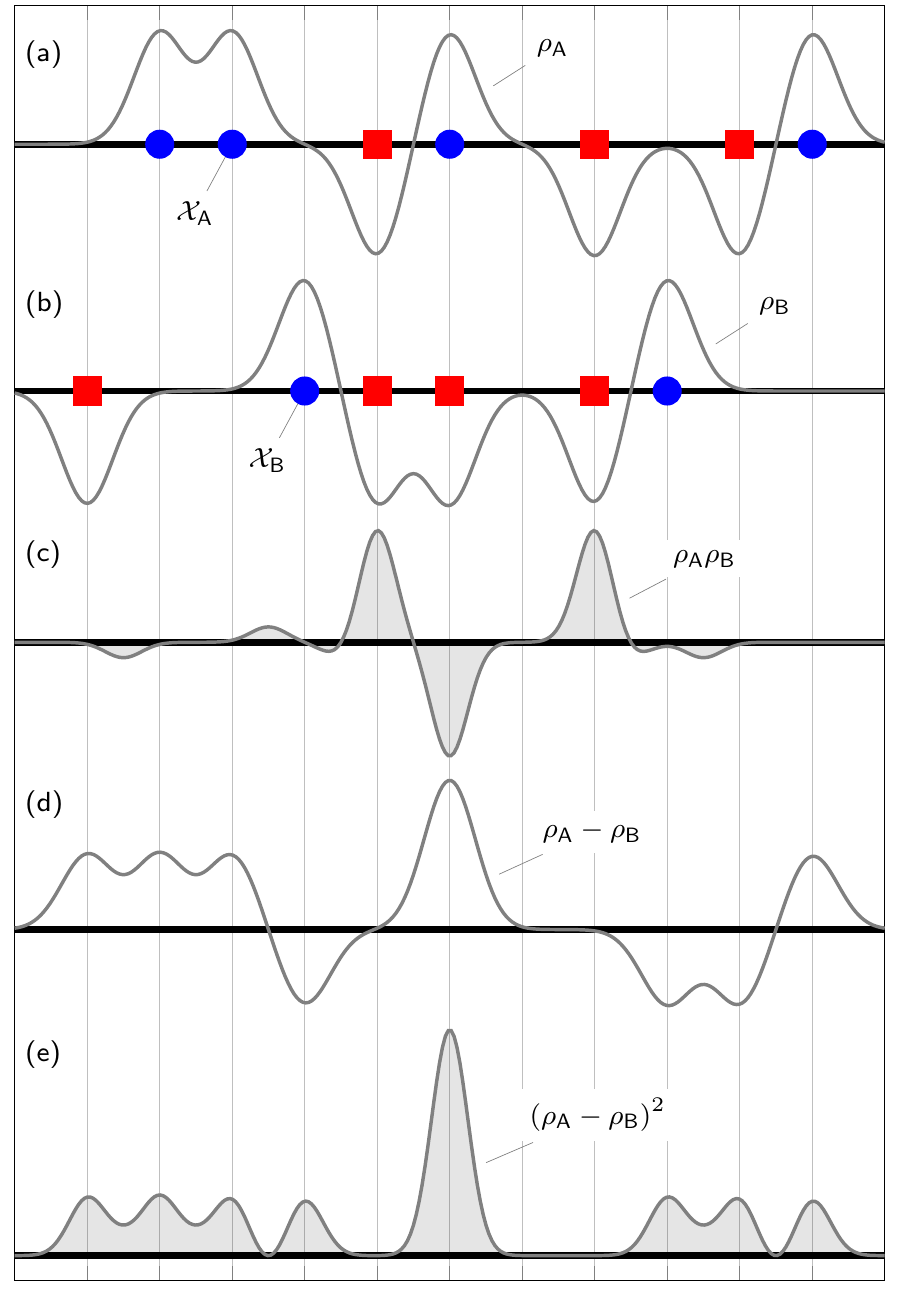}
\caption{One-dimensional schematic of the Gaussian Integral Inner Product and implied distance. Subfigures (a) and (b) respectively show atomic configurations A and B, each of which contain atoms of two types (indicated by circles and squares), and the atomic density functions $\rho_{\mathcal{X}_{\text{B}}}$ and $\rho_{\mathcal{X}_{\text{B}}}$ formed by placing weighted Gaussians at each atom site. This schematic illustrates the sign-species convention for Gaussian weights, where one atom type has positive weights, and the other atom type has negative weights. Subfigure (c) shows the product of the atomic density functions A and B, with the integrated shaded area being equal to $\left\langle \mathcal{X}_{\text{A}}, \mathcal{X}_{\text{B}} \right\rangle$. Subfigure (d) shows the distance in the two atomic density functions, the square of which is shown in (e). The integrated shaded area in (e) is equal to the squared distance between atomic configurations A and B.}
\label{fig:explainGiip}
\end{figure}

The Gaussian Integral Inner Product is inspired by the Smooth Overlap of Atomic Positions (SOAP) formulation \autocite{bartok2013representing}, centering a Gaussian function on each atom to create a continuous atomic density function, and then comparing atomic density functions as proxies for their corresponding (discrete) atomic environments. The various concepts in this section are illustrated in simplified form in Fig.~\ref{fig:explainGiip}.

We preliminarily establish the integral inner product of two functions in three-dimensional real space and the norm implied by that integral inner product (where $a$ and $b$ are arbitrary functions such that the integrals below are convergent) \autocite{schmidt1908auflosung}.
\begin{equation}
\left\langle a,b \right\rangle = \int_{\mathbb{R}^3} a(\mathbf{x}') \cdot b(\mathbf{x}')\ d\mathbf{x}'
\label{eq:innerProduct}
\end{equation}
\begin{equation}
||a|| = \sqrt{\left\langle a,a \right\rangle}
\label{eq:norm}
\end{equation}
We construct a three-dimensional Gaussian function $G$ centered at zero with standard deviation $\sigma$ and normalized so that the integral inner product norm (Eqn.~\eqref{eq:norm}) of $G$ is unity.
\begin{equation}
G_{\sigma}(\mathbf{x}) = \exp\left[ -|\mathbf{x}|^2 / (2 \sigma^2) \right]
/ (\pi^{3/4} \sigma^{3/2})
\label{eq:gaussian}
\end{equation}
We will use $G$ to construct atomic density functions from atomic configurations, illustrated in Fig.~\ref{fig:explainGiip}(a,b) for a simple one-dimensional case. Let $\mathcal{X}$ be an atomic configuration, consisting of a set of vectors pointing to atomic positions. We define an atomic density function $\rho$ by placing a weighted, shifted $G$ gaussian function at each atomic position. We do not constrain the weights ($w$) at this time, but they will be used to both encode species information and to handle behavior at a cutoff radius. When the weight of any given atom is zero, that atom effectively vanishes from the atomic configuration. Note that in addition to each atom having its own weight, each atom can also have its own $\sigma$, though in practice we generally use a constant $\sigma$ for all atoms in a configuration.
\begin{equation}
\rho_\mathcal{X}(\mathbf{x}) = \sum_{\mathbf{x'}\in\mathcal{X}} w_\mathbf{x'} G_{\sigma_\mathbf{x'}}(\mathbf{x} - \mathbf{x}')
\label{eq:atomicDensityFunction}
\end{equation}
Combining Eqns~\eqref{eq:innerProduct} and \eqref{eq:atomicDensityFunction}, we define the Gaussian Integral Inner Product (GIIP) between two atomic configurations ($\mathcal{X}^\alpha$ and $\mathcal{X}^\beta$):
\begin{equation}
\left\langle \mathcal{X}^\alpha,\mathcal{X}^\beta \right\rangle =
\left\langle \rho_{\mathcal{X}^\alpha},\rho_{\mathcal{X}^\beta} \right\rangle
\label{eq:giip}
\end{equation}
illustrated schematically in Fig.~\ref{fig:explainGiip}(c). We can also define the GIIP distance between those configurations using Eqn~\eqref{eq:norm} and the distributive property of the integral inner product:
\begin{align}
d_{\text{g}}\left(\mathcal{X}^\alpha,\mathcal{X}^\beta\right)^2 
&= 
||\rho_{\mathcal{X}^\alpha} - \rho_{\mathcal{X}^\beta}||^2 
\label{eq:giipDistanceA} \\
&=
\left\langle \rho_{\mathcal{X}^\alpha} - \rho_{\mathcal{X}^\beta}, \rho_{\mathcal{X}^\alpha} - \rho_{\mathcal{X}^\beta} \right\rangle 
\label{eq:giipDistanceB} \\
&=
\left\langle \rho_{\mathcal{X}^\alpha},\rho_{\mathcal{X}^\alpha} \right\rangle 
+ \left\langle \rho_{\mathcal{X}^\beta},\rho_{\mathcal{X}^\beta} \right\rangle 
- 2\cdot\left\langle \rho_{\mathcal{X}^\alpha},\rho_{\mathcal{X}^\beta} \right\rangle 
\label{eq:giipDistanceC} \\
&= 
\left\langle \mathcal{X}^\alpha,\mathcal{X}^\alpha \right\rangle 
+ \left\langle \mathcal{X}^\beta,\mathcal{X}^\beta \right\rangle 
- 2\cdot\left\langle \mathcal{X}^\alpha,\mathcal{X}^\beta \right\rangle
\label{eq:giipDistanceD}\end{align}
illustrated in Fig.~\ref{fig:explainGiip}(d-e). The GIIP distance defined above is immediately usable as an orientation-sensitive measurement between atomic configurations. It is possible to render the GIIP distance orientation-invariant by finding the minimum distance over all possible orientations:
\begin{equation}
d_{\text{g,inv}}\left(\mathcal{X}^\alpha,\mathcal{X}^\beta\right)^2  = \min_{\mathbf{R} \in \text{O(3)}}\left|\mathcal{X}^\alpha-\mathbf{R}\mathcal{X}^\beta\right|^2
\label{eq:giipDistanceOrientationInvariant}
\end{equation}
where O(3) is the three-dimensional orthogonal group (comprising rotations and rotoinversions) and $\mathbf{R}\mathcal{X}$ is a specific rigid-body rotation or rotoinversion of the atomic environment; note that in cases where chirality must be preserved, this minimization can be done over SO(3) instead.

The atomic weights in the GIIP formulation serve three purposes: to exclude atoms from analysis (e.g. outside of a cutoff radius), to enforce continuity or smoothness with respect to perturbation of atomic positions near (for example) a cutoff radius, and to encode species information. Atoms may be excluded from analysis by setting their weights to zero, eliminating their effect on the atomic density function (Eqn.~\eqref{eq:atomicDensityFunction}). This might be desirable when excluding extraneous information (perhaps, for example, ignoring the positions of hydrogen atoms in a polymer), or when excluding regions from analysis, as in the case of atoms outside of a cutoff radius. In cases where continuity or smoothness at a cutoff radius is desired, weights can be thought of as being the product of a function of $|\mathbf{x}|$ and of a function of species:
\begin{equation}
w_{\mathbf{x}} = f(|\mathbf{x}|)g(\text{species}(\mathbf{x}))
\label{eq:weights}
\end{equation}
By having $f$ continuously or smoothly vanish as it approaches the cutoff radius from below, continuity and smoothness of the GIIP distance with respect to atomic position are enforced across that cutoff radius. We suggest two strategies for encoding species information into weights. The first strategy, which we refer to as ``sign-species,'' is effective for binary systems, and consists of setting weights corresponding to one atom type to positive values, and weights corresponding to the second atom type to negative values. The second strategy, which we refer to as ``vector-species,'' is effective for binary and higher-order systems and consists of setting weights for an $n$-ary system to be $n$-vectors, where the weight vector for an atom of type $i$ will consist entirely of zeroes except for its $i$'th coordinate. This approach essentially reduces to computing $n$ separate GIIP distances, each excluding all but a single atom type, and then summing to obtain a single-valued distance value. Note that this summation must occur inside of the optimization loop when computing the orientation-invariant GIIP distance in Eqn.~\eqref{eq:giipDistanceOrientationInvariant}.

The GIIP formalism is equally applicable in one, two, or three dimensions and analytically tractable by substituting Eqns~\eqref{eq:gaussian} and \eqref{eq:atomicDensityFunction} into \eqref{eq:giip} and integrating \eqref{eq:innerProduct}:
\begin{equation}
\left\langle \mathcal{X}^\alpha,\mathcal{X}^\beta \right\rangle =
2\sqrt{2} \sum_{\mathbf{x}^\alpha \in\mathcal{X}^\alpha}  \sum_{\mathbf{x}^\beta\in\mathcal{X}^\beta}
w_{\mathbf{x}^\alpha} w_{\mathbf{x}^\beta} \left(\frac{\sigma_{\mathbf{x}^\alpha} \sigma_{\mathbf{x}^\beta}}{\sigma_{\mathbf{x}^\alpha}^2 + \sigma_{\mathbf{x}^\beta}^2}\right)^{3/2}
\exp\left[ -|\mathbf{x}^\alpha - \mathbf{x}^\beta|^2 / (2\sigma_{\mathbf{x}^\alpha}^2 + 2\sigma_{\mathbf{x}^\beta}^2) \right]
\label{eq:giipEvaluated}
\end{equation}
which, in the case where $\sigma_{\mathbf{x}}$ is constant for all atoms, simplifies to:
\begin{equation}
\left\langle \mathcal{X}^\alpha,\mathcal{X}^\beta \right\rangle =
\sum_{\mathbf{x}^\alpha \in\mathcal{X}^\alpha}  
\sum_{\mathbf{x}^\beta\in\mathcal{X}^\beta}
w_{\mathbf{x}^\alpha} w_{\mathbf{x}^\beta}
\exp\left[ -|\mathbf{x}^\alpha - \mathbf{x}^\beta|^2 / (4\sigma^2) \right]
\label{eq:giipEvaluatedSimplified}
\end{equation}

We implemented Eqns.~\eqref{eq:giipDistanceD}, \eqref{eq:giipDistanceOrientationInvariant}, \eqref{eq:weights}, and \eqref{eq:giipEvaluatedSimplified} in a Python library based on the popular PyTorch library \autocite{NEURIPS2019_9015} with both thread-based and GPU-based parallelism, and executed the computations shown here on computers ranging from a laptop to a large institutional cluster. The computations were accelerated by condensing all atomic positions into tensors of size $n_{\text{configurations}} \times \max\left\lbrace n_{\text{atoms}}\right\rbrace \times n_{\text{dimensions}}$, and all weights into tensors of size $n_{\text{configurations}} \times \max\left\lbrace n_{\text{atoms}}\right\rbrace$ (where $\max\left\lbrace n_{\text{atoms}}\right\rbrace$ represents the maximum number of atoms found in any configuration). In neighborhoods with fewer atoms than $\max\left\lbrace n_{\text{atoms}}\right\rbrace$, weights corresponding to non-existent atoms are set to zero. We used the hyperspherical-coverings library \autocite{larsen2017improved} to sample orientation space in evaluating Eqn.~\eqref{eq:giipDistanceOrientationInvariant}.

In the analysis of the unary two-dimensional crystal shown here, we used the sum of the weights in an LAE as the coordination number of the central atom of the LAE. For the three-dimensional NiNb glass, we used the sign-species formalism, so the sum of weights in the LAE was linearly related to the composition of the nearest neighbors of the central atom.

For the two-dimensional crystal, we used a uniform weight of 1 for atoms less than 1.3$\sigma_{\text{LJ}}$ (where $\sigma_{\text{LJ}}$ is the length-scale of the Lennard-Jones formalism) from the central atom, and had the weight smoothly drop to zero (using a cubic spline) for atoms between $1.3\sigma_{\text{LJ}}$ and $1.75\sigma_{\text{LJ}}$ from the central atom; in our GIIP calculations we used $\sigma=0.5\sigma_{\text{LJ}}$ and sampled orientation space with a resolution of one degree. 

For the three-dimensional metallic glass we used a uniform weight of +1/-1 for Ni/Nb less than 3 \AA{} from the central atom, and had the weight smoothly drop to zero between 3 \AA{} and 4 \AA{}. In our GIIP calculations we used $\sigma=1.0$ and sampled three-dimensional orientation space first with a coarse resolution of 5 degrees and then fine resolution of 1 degree around the minimum of the coarse search.

\subsection*{Cluster Analysis}
Agglomerative clustering assigns datapoints to a prescribed number of classes, where each class contains datapoints that are similar by some measure. Here it enables us to break the atomic configurations into classes; for example, in a crystalline material we might establish classes as:
a)	atoms in a perfect lattice,
b)	atoms bordering a single vacancy,
c)	interstitial atoms,
and so forth. In other words, agglomerative clustering provides a \textit{discrete} parameterization of local structure. There might be some variation among the configurations in each class (for example, due to thermal vibrations or strain fields), but for the classification to be useful the configurations within each class must be similar enough to behave similarly, with the threshold of similarity ultimately being a user decision. There are many potential agglomerative clustering algorithms available in literature; in this work we used a patient diameter-minimizing criterion \autocite{lance1967general} but it is possible that other clustering algorithms would yield superior results. We refer the reader to any textbook on data mining for conceptual details, and to the scipy.cluster package \autocite{2020SciPy-NMeth} for an accessible implementation.

\subsection*{Diffusion Maps}
Dimensionality reduction algorithms are used for computing a set of data-driven latent coordinates. Here, we used Diffusion Maps \autocite{coifman2006diffusion}, a manifold learning scheme. Diffusion maps offer a reparametrization of the original data by revealing its \textit{intrinsic} geometry. Below we give a short description of the algorithm.

The diffusion maps algorithm is applied to a given data set $\textbf{X} =\{\vect{x}_i\}_{i}^n$ sampled from a manifold $\mathcal{M}$, where we assume $\vect{x}_i \in \mathbb{R}^m$ first constructs a random walk on the data. This random walk is estimated based on the local \textit{similarity} of the sampled data points. The \textit{similarity} measure is computed in terms of a kernel, for example the Gaussian kernel, defined as
\begin{equation}
    W(\vect{x}_i,\vect{x}_j) = \exp\bigg( \frac{-||\vect{x}_i - \vect{x}_j ||^2}{2\varepsilon^2} \bigg).
\end{equation}
where $||\cdot||$ is an appropriate norm; in our case the GIIP distance. The hyper-parameter $\varepsilon$ is the kernel's scale. In our work we selected $\varepsilon$ by trial-and-error; for the two-dimensional crystal we used $\varepsilon=1.0$ atoms of GIIP distance and for the three-dimensional metallic glass we used $\varepsilon=1.5$ atoms of GIIP distance.

To recover a parametrization of the data set regardless of its sampling density a normalization on $\textbf{W}$ is computed,
\begin{equation}
    \widetilde{\mathbf{W}} = \mathbf{P}^{-1}\mathbf{W}\mathbf{P}^{-1}.
\end{equation}
where the diagonal matrix $\textbf{P} \in \mathbb{R}^{m \times m}$ is computed by,
\begin{equation}
    P_{ii}  = \sum_{j=1}^nW_{jj}
\end{equation}

A second normalization is then applied to $\widetilde{\textbf{W}}$ to construct a Markovian matrix $\textbf{M}$,
\begin{equation}
    M(\vect{x}_i,\vect{x}_j) = \frac{\widetilde{W}(\vect{x}_i,\vect{x}_j)}{\sum_{j=1}^n \widetilde{W}(\vect{x}_i,\vect{x}_j)}
\end{equation}
Computing the eigendecomposition of $\textbf{M}$ gives a set of eigenvalues $\lambda_i$ and eigenvectors $\vect{\phi}_i$
\begin{equation}
    \textbf{M}\vect{\phi}_i  = \lambda_i\vect{\phi}_i
\end{equation}
The eigenvectors of $\textbf{M}$ are the data-driven coordinates that offer a reparametrization of $\textbf{X}$. However,a selection of the eigenvectors that span independent directions is needed. Those independent eigenvectors are called non-harmonics, and we refer the reader to \autocite{dsilva2018parsimonious} and also in the SI of \autocite{evangelou2022double} for a more detailed discussion of non-harmonic eigenvectors. If the number of those independent/non-harmonic eigenvectors is less than original dimensions of $\textbf{X}$ then we claim the diffusion maps algorithm achieved dimensionality reduction.

To obtain the diffusion coordinates for points $\vect{x}_{new} \notin \textbf{X}$ without recomputing the entire diffusion map the Nystr\"om Extension \autocite{fernandez2015diffusion,Nystrom_Extension} can be used. Nystr\"om Extension computes the new coordinates based on a weighted average,
\begin{equation}
\label{eq:Nystrom}
    \phi_i(\vect{x}_{new}) = \frac{1}{\lambda_i}\sum_{j=1}^n \widetilde{M}(\vect{x}_{new},\vect{x}_j)\phi_i(\vect{x}_j),
\end{equation}
where $\vect{\phi}_i(\vect{x}_j)$ denotes the $j$-th component of the $i$-th eigenvector and $\vect{\phi}_i(\vect{x}_{new}$) is the estimated $i$-coordinate for the \textit{out of sample} data point $\vect{x}_{new}$. To compute the kernel $\widetilde{\vect{M}}$ the same normalizations and the same scale parameter $\varepsilon$ used for diffusion maps are needed.

Finally, we note that the orientation-invariant GIIP distance does not satisfy the triangle inequality, with the consequence that the kernel matrix is only approximately symmetric positive definite. In our metallic glass dataset, the first hundred eigenvalues of the kernel matrix were positive, but several relatively small (roughly 1/2 of a percent of the largest eigenvalue in magnitude) negative eigenvalues appeared at higher orders. This places use of the GIIP distance outside of portions of theory for diffusion maps, but we deem our kernel matrix to be ``close enough'' to symmetric positive definiteness, and our results to be sensical enough to accept this, particularly in view of the absurdity of using more than one hundred diffusion coordinates in this dimensionality reduction exercise.

\subsection*{Sample construction}
The two-dimensional crystal was generated using the LAMMPS molecular dynamics package \autocite{LAMMPS}. We generated 20,000 atoms arranged in a hexagonal lattice with vacuum boundaries, interacting with a standard 12-6 Lennard-Jones interatomic potential with unit mass, distance, and energy terms, and a cutoff at $2.5\sigma_{\text{LJ}}$. We then introduced defects into the crystal by randomly selecting an atom and one of its six close-packed nearest neighbors, and then deleting the half-plane of atoms in the direction of that neighbor. This was repeated ten times. After thus modifying the sample, the system was equilibrated in the NVE ensemble with a timestep of 0.003 for 30,000 time steps. This resulted in a system with roughly ten defects consisting of both dislocations and vacancies.

The four three-dimensional metallic glass samples were also generated using LAMMPS using an embedded atom method potential \autocite{zhang2016experimental} constructed for NiNb glasses. For each sample, we generated 13500 atoms in an FCC lattice (intial $a=4$\AA), randomly assigning approximately half to be Ni and half to be Nb. We set the initial velocity of the atoms consistent with a temperature of 2500 Kelvin and then equilibrated the simulation in the NPT ensemble at a temperature of 2000 Kelvin and a pressure of 1 bar, with a timestep of 0.001 picoseconds for 10000 steps. Finally, we quenched the simulation in NPT mode from 2000 Kelvin to 300 Kelvin at quench rates for the four samples were $1.7\times 10^{13}$, $3.4\times 10^{12}$, $1.7\times 10^{12}$, and $1.7\times 10^{11}$ K/s, resulting in a glassy structure. These quench rates are too high to be experimentally relevant; however, high quench rates encouraged formation of a wide range of energetically unfavorable configurations for the manifold learning algorithm to learn.

\subsection*{Acknowledgments}
Thanks to Ian Winter, Mark Wilson, and N. Scott Bobbitt for enlightening conversations. 

This work was funded by the President Harry S. Truman Fellowship in National Security Science and Engineering and LDRD programs at Sandia National Laboratories. The JHU authors gratefully acknowledge support by subcontract under PO\#2221045 from Sandia National Laboratories.

\textit{This article has been authored by an employee of National Technology \& Engineering Solutions of Sandia, LLC under Contract No. DE-NA0003525 with the U.S. Department of Energy (DOE). The employee owns all right, title and interest in and to the article and is solely responsible for its contents. The United States Government retains and the publisher, by accepting the article for publication, acknowledges that the United States Government retains a non-exclusive, paid-up, irrevocable, world-wide license to publish or reproduce the published form of this article or allow others to do so, for United States Government purposes. The DOE will provide public access to these results of federally sponsored research in accordance with the DOE Public Access Plan https://www.energy.gov/downloads/doe-public-access-plan. This paper describes objective technical results and analysis. Any subjective views or opinions that might be expressed in the paper do not necessarily represent the views of the U.S. Department of Energy or the United States Government. (SAND2022-15908 O)}

\printbibliography

\end{document}